\documentclass[nofootinbib,notitlepage,superscriptaddress,10pt,aps,prd,onecolumn]{revtex4-1}
\usepackage[utf8]{inputenc}
\usepackage{amsmath,mathtools}

\usepackage{tensor}
\usepackage{amssymb}
\usepackage{graphicx}
\usepackage{physics}
\usepackage{hyperref}
\usepackage{natbib}
\usepackage{lipsum}  
\usepackage{bigints}
\usepackage{verbatim}
\newcommand{\lerisq}[1]{\left[#1\right]}
\newcommand{\leri}[1]{\left(#1\right)}

\begin{document}
\title{The role of longitudinal polarizations in Horndeski and macroscopic gravity: Introducing gravitational plasmas}
\author{Fabio Moretti}
\email{fabio.moretti@uniroma1.it}
\affiliation{Physics Department, ``Sapienza'' University of Rome, P.le Aldo Moro 5, 00185 (Roma), Italy}
\author{Flavio Bombacigno}
\email{flavio2.bombacigno@uv.es}
\affiliation{Departament de F\'{i}sica Teòrica and IFIC, Centro Mixto Universitat de València - CSIC, Universitat de València, Burjassot 46100, València, Spain}
\author{Giovanni Montani}
\email{giovanni.montani@enea.it}
\affiliation{ENEA, Fusion and Nuclear Safety Department, C. R. Frascati,
	Via E. Fermi 45, 00044 Frascati (Roma), Italy}
\affiliation{Physics Department, ``Sapienza'' University of Rome, P.le Aldo Moro 5, 00185 (Roma), Italy}

\begin{abstract}
We discuss some general and relevant features of longitudinal gravitational modes in Horndeski gravity and their interaction with matter media. Adopting a gauge-invariant formulation, we clarify how massive scalar and vector fields can induce additional transverse and longitudinal excitations, resulting in breathing, vector, and longitudinal polarizations. 
We review, then, the interaction of standard gravitational waves with a molecular medium, outlining the emergence of effective massive gravitons, induced by the net quadrupole moment due to molecule deformation. Finally, we investigate the interaction of the massive mode in Horndeski gravity with a noncollisional medium, showing that Landau damping phenomenon can occur in the gravitational sector as well. That allows us to introduce the concept of ``gravitational plasma'', where inertial forces associated with the background field play the role of cold ions in electromagnetic plasma.
\end{abstract}
\maketitle
\section{Introduction}
 The recent observation of gravitational waves emitted from black hole and neutron star coalescence~\cite{Abbott:2017vtc,LIGOScientific:2016aoc,LIGOScientific:2016sjg,LIGOScientific:2017zic,LIGOScientific:2017vox,LIGOScientific:2018mvr,LIGOScientific:2020ibl,LIGOScientific:2021djp,LIGOScientific:2021usb}, besides offering an important confirmation to weak field predictions of general relativity~\cite{Maggiore:2007ulw,Maggiore:2018sht}, opens new perspectives in the investigation of astrophysical and cosmological phenomena~\cite{Chatziioannou:2012rf,Healy:2011ef,Berti:2013gfa,Shibata:2013pra,Bambi:2015kza,Isi:2015cva,Cardoso:2016oxy,Yunes:2016jcc,LIGOScientific:2016lio,LIGOScientific:2016vpg,Ezquiaga:2017ekz,LIGOScientific:2018dkp,Crisostomi:2017lbg,Calcagni:2019ngc,LIGOScientific:2019fpa,LIGOScientific:2020tif}. It is clear that the choice of the Einstein--Hilbert action for the gravitational dynamics corresponds just to a criterion of simplicity, ensuring the second-order differential character of the field equations (see Lovelock \mbox{theorem~\cite{Lovelock:1971yv,Lovelock:1972vz}}).

The study of more general formulations was originally motivated by the expectation that the singularities emerging in general relativity, such as the initial big-bang singularity or the diverging curvature characterizing the center of Schwarzschild black holes, could be solved in the presence of different geometrical terms in the Lagrangian; for instance, with higher-order contributions~\cite{Jackiw:2003pm,Flanagan:2003iw,Nojiri:2005jg} or further contractions~\cite{BeltranJimenez:2017doy,Afonso:2018mxn,Afonso:2018bpv,Delhom:2019wcm,Delhom:2019zrb,Iosifidis:2020dck,Bombacigno:2021bpk}. Higher-order terms were also prompted by the request of recovering renormalizability at the quantum level, and quadratic curvature invariants were shown to properly deal with the problem~\cite{Stelle:1976gc,Benedetti:2009rx,Modesto:2011kw,Lauscher:2001rz,Salvio:2018crh}. On the other hand, since their inclusion would evidently have implied field equations of a higher order than two, the issue of possible Ostrogradsky instabilities led Horndeski to define the most general scalar tensor theory of gravity, endowed with an additional scalar field beside the metric, devoid of pathologies and preserving the second-order nature of the equations of motion~\cite{Horndeski:1974wa}. Recently, the motivation for extended theories of gravity has been also driven by the request to provide a satisfactory explanation for the Universe's dark components, e.g., dark matter and dark energy~\cite{Cognola:2007zu,Nojiri:2010wj,Capozziello:2011et,Bamba:2012cp,Capozziello:2012qt,Capozziello:2012ny,Tamanini:2013ltp,Bahamonde:2017ize,Rosa:2017jld,Nojiri:2017ncd}. 

It is important to stress that reformulations of the Einstein--Hilbert action can be carried on not only in the direction of a revised Lagrangian, but also according to a metric-affine perspective, where the affine connection is assumed to be an independent geometrical object with respect to the metric, and whose form has to be dynamically derived by evaluating its very equation of motion~\cite{Afonso:2017bxr,Iosifidis:2018jwu,BeltranJimenez:2019acz,Iosifidis:2019fsh}. These alternative approaches are usually enriched with a nontrivial geometric structure, where curvature is accompanied by torsion and nonmetricity~\cite{Latorre:2017uve,Capozziello:2007tj,Olmo2011,Afonso:2017bxr,Iosifidis:2018zjj,Delhom:2019yeo}---they turned out to be capable of tackling singularity problems~\cite{Olmo:2015axa,Bambi:2015zch,Olmo:2016fuc,Menchon:2017qed,Nascimento:2018sir,Bejarano:2019zco,Guerrero:2021ues,Guerrero:2021avm,Olmo:2020fnk} or cosmological observations~\cite{Bombacigno2016,Bombacigno:2018tyw,Bombacigno:2021bpk,Benisty:2021laq,BeltranJimenez:2017uwv} from an original standpoint.

Clearly, such new formulations must be reconciled with observational and experimental constraints, and deviations from general relativity in the Solar System are very strict~\cite{Berry:2011pb,Iorio:2007rk}. Nonetheless, increasing interest emerged for the possibility to observe modifications in the detected gravitational waves, such as additional polarization modes~\cite{LIGOScientific:2017ous,LIGOScientific:2018czr,Callister:2017ocg,Andriot:2017oaz,Sagunski:2017nzb,Liang2017,Hou:2017bqj} or anomalous dispersive and dissipative behaviors~\cite{Odintsov:2021kup}, hopefully originating from extended dynamical paradigms. 

Here, we will concentrate our attention on the morphology and propagation of gravitational waves in the context of Horndeski gravity, in the broader sense of scalar and vector couplings~\cite{Horndeski:1976gi,Tasinato:2014eka,Allys:2015sht,Allys:2016jaq,Rodriguez:2017ckc,GallegoCadavid:2019zke,Allys:2016kbq,GallegoCadavid:2020dho,GallegoCadavid:2021ljh,Rodriguez:2017wkg,Gomez:2019tbj,Garnica:2021fuu,Heisenberg:2014rta,Heisenberg:2016eld,Heisenberg:2018acv,BeltranJimenez:2013btb}. In particular, in this framework, we will analyze the polarizations induced in the geodesic deviation equation by the massive scalar or vector modes, discussing to some extent the limit of vanishing mass. Furthermore, we will study the interaction of gravitational waves with a material medium, having the morphology of a molecular array or plasmalike features. In this respect, we will introduce some new subtle concepts such as ``Macroscopic Gravity Theory''~\cite{Montani:2018iqd} or ``Gravitational Plasma''~\cite{Moretti:2020kpp}.

In more detail, regarding the investigation of the gravitational modes in Horndeski gravity, we review and partially extend the gauge-invariant formulation of the linear theory~\cite{Moretti2019} in order to unequivocally identify the physical degrees of freedom. Specifically, we arrive to a clear demonstration that scalar and vector fields are responsible for a superposition of transverse and longitudinal excitations, whose properties depend crucially on the massive nature of the additional polarizations. This overview of the gauge-invariant formulation fixes a specific and quite general phenomenological signature for Horndeski theories, and offers remarkable suggestions about the device that could optimize the detection of such anomalous modes.

 In investigating the interaction of gravitational waves with matter media, we first review some well-known results about the attenuation of the signal due, for instance, to the dissipation properties of the traveled medium~\cite{Hawking:1966qi,Madore}, to the redshift in an expanding Universe~\cite{1978SvA....22..528Z,Weinberg:2003ur,Flauger:2017ged}, or to the interaction with a cosmological neutrino \mbox{background~\cite{Lattanzi:2005xb,Lattanzi:2010gn,Benini:2010zz}.} Then, we face the question concerning the propagation of waves in a molecular or a plasmalike medium. In the case of matter medium characterized by molecular substructures~\cite{Montani:2018iqd}, as it occurs in astrophysical systems such as galaxies, we show that standard gravity acquires an effective massive behavior, resulting in five degrees of freedom. In this scenario, anomalous, massivelike polarizations appear and their features are discussed in some detail. This phenomenon is very similar to what happens to electromagnetic waves interacting with a plasma: the photon acquires an effective mass and a longitudinal polarization is present inside the plasma~\cite{Mendonca:2000tk,PhysRevE.49.3520}. Analogously, the interaction of the graviton in a molecular medium can be restated in terms of effective massive states, which propagate with subluminal velocity.

Then, we consider a different medium, made up of free-falling particles in the gravitational field, whose perturbations are now described by Horndeski gravity. We speak of ``Gravitational Plasma'' when we can choose a local inertial frame, somehow analogous to a ``neutralizing field'', which in real plasma is typically associated to cold ions. The idea is that massive modes of the Horndeski formulation can interact with such a system analogously to longitudinal electromagnetic waves in real plasmas of ions and electrons. In other words, we attribute the local gravitational field with the neutralizing role of inertial forces and we analyze the propagation of the massive mode in a self-consistent way, as for Langmuir modes in the electromagnetic case~\cite{pitaevskii2012physical}. The settling of the Landau damping for the massive mode is then confirmed via a kinetic description of the medium-wave interaction. This result establishes an important parallelism between electromagnetism in plasma and Horndeski gravity for free-falling particles, i.e., noncollisional ``astrophysical'' or ``cosmological'' plasma.

The present review offers an interesting tool to deepen the comprehension of the role of longitudinal gravity waves, as predicted by the extended formulation of gravity, can play when interacting with physical media of different typologies. The review is structured as follows: In Section~\ref{sec2}, we discuss the propagation of massive scalar and vector fields in Horndeski theories, outlining the different kinds of polarizations carried out by these additional degrees of freedom and the stresses induced on test particles via the geodesic deviation equation. In Section~\ref{sec3}, we review some known results about the interaction of gravitational waves with matter and we eventually establish formal and phenomenological analogies between electromagnetic plasmas and linearized gravitational waves in matter. We firstly introduce the concept of the effective massive graviton in the presence of a molecular stress energy tensor; then, we discuss the idea of the gravitational Landau damping for longitudinal scalar modes. Finally, in Section~\ref{sec4}, conclusions are drawn.

\section{Longitudinal Degrees of Freedom and Gauge-Invariant Technique}\label{sec2}

 Let us focus our analysis on metric perturbations over Minkowski spacetime, i.e.,
\begin{equation}
g_{\mu\nu}=\eta_{\mu\nu}+h_{\mu\nu},
\label{pert metr}
\end{equation}
with $\eta_{\mu\nu}=diag(-1,1,1,1)$ and where we are assuming there exists some reference frame in which $|h_{\mu\nu}|\ll 1$ holds. It follows, then, that in order for $g^{\mu\rho}g_{\rho\nu}=\delta^\mu_\nu+\mathcal{O}(h^2)$ to remain consistent, the inverse metric must retain the form 
\begin{equation}
g^{\mu\nu}=\eta^{\mu\nu}-h^{\mu\nu}.
\label{pert metr inv}
\end{equation}

It is worth noting that in the context of modified theories of gravity, the set of ground states actually available can be significantly larger than the ``trivial'' vacuum Minkowski solution. The enhanced dynamics featuring these theoretical settings, indeed, can in principle sustain additional configurations with respect to the globally flat state described by $\eta_{\mu\nu}$.
However, since the current detection of gravitational waves by means of ground-based interferometers is ultimately not affected in an appreciable way by background curvature (for truly local effects, such as Newtonian noise, see~\cite{2018PhRvD..97f2003F}), we can safely restrict our treatment to gravitational field excitations traveling on Minkowksi background. As we are primarily interested in determining the number and type of additional polarizations, we can as a matter of fact ignore any background configuration featured by a nonvanishing curvature, which we expect could instead affect evolutionary properties of the wave perturbation, such as amplitude and frequency. We refer, for instance, to~\cite{Naf:2008sf,Bernabeu:2011if,Ozer:2021qjb}, where the effect of a cosmological constant is discussed in general relativity and Brans--Dicke theories, outlining effects eventually measurable with the pulsar timing array \mbox{technique~\cite{2010CQGra..27h4013H,2011ASSP...21..229H,Lentati:2015qwp},} or~\cite{Ananda:2006af,Caprini:2015zlo,LIGOScientific:2017adf,Caprini:2018mtu,BICEP2:2018kqh,Maggiore:2018sht,Flauger:2017ged} where propagation on cosmological background are also considered. 
Given such premises, we decompose the metric perturbation $h_{\mu\nu}$, which we assume to be vanishing for $r\to\infty$ (asymptotically flatness is recovered), in its irreducible components (see~\cite{Weinberg:2008zzc,Flanagan:2005yc}):
\begin{equation}
\begin{split}
h_{00}&=2\phi ,\\
h_{0i}&=\beta_i+\partial_i \gamma ,\\
h_{ij}&=h^{TT}_{ij} + \dfrac{1}{3}H\delta_{ij}+\partial_{(i}\epsilon_{j)}+\left(\partial_i\partial_j-\dfrac{1}{3}\delta_{ij}\bigtriangleup\right)\lambda,
\end{split}
\label{splitting metric}
\end{equation}
where $\delta_{ij}$ is the Kronecker delta, $\bigtriangleup\equiv\partial_i\partial^i$ is the flat Laplacian operator, and symmetrization is defined as $A_{(ij)}\equiv \frac{1}{2}(A_{ij}+A_{ji})$. It is then evident that the splitting \eqref{splitting metric} is somehow redundant, given that we defined $16$ new functions for describing a symmetric tensor of rank two \footnote{When we work within the linearized theory, we can consider in all respects the perturbation $h_{\mu\nu}$ as a dynamical tensor field living on a Minkowski spacetime---see, for instance, the discussion in Ch.~2 of~\cite{Maggiore:2007ulw}.}. In order to restore the proper number of components, it is thus necessary to introduce the set of constraints

\begin{equation}
\begin{split}
\partial^i \beta_i&=0 \\
\partial^i h^{TT}_{ij}&=0 \\
\eta^{ij}h^{TT}_{ij}&=0\\
\partial^i \epsilon_i &=0,
\end{split}
\label{constraint metric}
\end{equation}
which, by fixing six additional conditions, allows us to recover the ten independent elements carried by $h_{\mu\nu}$. In general, when matter is also included in the analysis, the metric perturbation contains a combination of gauge and physical degrees of freedom, where the latter encompass, in turn, radiative and nonradiative fields. The first ones represent self-sustained perturbations, which can propagate freely in vacuum, while the second ones are only excited in the presence of matter sources. In general relativity, however, this crucial distinction is partially obscured when dealing with specific gauge choice, for instance, in the Lorentz gauge $\partial_\mu h^{\mu\nu}=0$, since in this case, the equation of motion can be simply rearranged in the form
\begin{equation}
    \Box h_{\mu\nu}=-2\kappa T_{\mu\nu},
    \label{equation gravitational wave}
\end{equation}
which seems to suggest that all the remaining six components of the metric perturbation could be radiative. Only in vacuum can the residual gauge freedom be further exploited to extract from $h_{\mu\nu}$ the two physical degrees of freedom truly propagating, i.e., the transverse and traceless states encoded in $h^{TT}_{ij}$, corresponding to the well-known cross ``$\times$'' and plus ``$+$'' polarizations. Therefore, given that it seems sensible to reformulate the problem without fixing a priori the system of coordinates---that is to say, in a gauge-invariant way, in order to obtain directly from the equations of the motion the dynamic properties of the theory. We stress, however, that such a statement has to be considered only at the linearized level, since in performing \eqref{pert metr}, we have already selected a special gauge. Let us consider the infinitesimal coordinate transformation 
\begin{equation}
    x'^\mu=x^\mu+\xi^\mu(x),
    \label{diffeo lin}
\end{equation}
which, once we take into account the general transformation rule for the metric
\begin{equation}
    g'_{\mu\nu}(x')=\frac{\partial x^\rho}{\partial x'^\mu}\frac{\partial x^\sigma}{\partial x'^\nu}g_{\rho\sigma}(x),
\end{equation}
leads to the gauge transformation for the perturbation $h_{\mu\nu}$:
\begin{equation}
    h'_{\mu\nu}(x')=h_{\mu\nu}(x)-2\partial_{(\mu}\xi_{\nu)}.
\end{equation}

It is therefore clear that for \eqref{pert metr} to remain 
consistent, i.e., $|h_{\mu\nu}|\ll 1$, we require \linebreak $\partial\xi\sim\mathcal{O}(h)$. In this respect, it is interesting to note that we are not compelled to demand for $\xi^\mu$ itself to be of the same order of $h_{\mu\nu}$ but only its derivative, which also guarantees that we could consistently solve \eqref{diffeo lin} for $x^\mu$. We introduce, then, the combinations \footnote{For all details concerning the properties and the asymptotic behavior of $h_{\mu\nu}$, $\xi$, and their decomposition, we remind the reader to~\cite{Flanagan:2005yc}.}
\begin{equation}
\begin{split}
\Phi &= -\phi + \dot{\gamma}-\dfrac{1}{2}\ddot{\lambda} \\
\Theta &= \dfrac{1}{3} \left ( H-\bigtriangleup \lambda\right)\\
\Xi_i &= \beta_i-\dfrac{1}{2}\dot{\epsilon}_i,
\end{split}
\label{eq15}
\end{equation}
where the dot denotes time derivative, which is easy to see to constitute, together with $h^{TT}_{ij}$, the correct number of gauge-invariant quantities. By other words, instead of dealing directly with the entire tensor $h_{\mu\nu}$, we can simply consider the six independent components carried by the two scalars $\Phi$ and $\Theta$, the divergenceless vector $\Xi_i$, and the transverse and traceless matrix $h^{TT}_{ij}$. Hence, we proceed by performing an equivalent decomposition on the stress energy tensor, i.e.,
\begin{equation}
\begin{split}
T_{00}&=\rho,\\
T_{0i}&=S_i+\partial_i S ,\\
T_{ij}&=\sigma_{ij}+P\delta_{ij}+\partial_{(i}\sigma_{j)} + \left( \partial_i\partial_j-\dfrac{1}{3}\delta_{ij}\bigtriangleup\right)\sigma,
\end{split}
\label{eq16}
\end{equation}
which are accompanied by an analogous set of constraints, namely,
\begin{equation}
\begin{split}
\partial_i S^i& =0 \\
\partial^i\sigma_{ij}&=0\\
\eta^{ij}\sigma_{ij}&=0\\
\partial_i \sigma^i&=0.
\end{split}
\label{eq17}
\end{equation}

Moreover, by taking into account the linearized conservation law $\partial_\mu T^{\mu\nu}=0$, we can derive an additional set of relations, i.e.,
\begin{equation}
\begin{split}
\bigtriangleup S &=\dot{\rho}, \\
\bigtriangleup \sigma &= -\dfrac{3}{2}P+\dfrac{3}{2}\dot{S}, \\
\bigtriangleup \sigma_i&=2 \dot{S}_i,
\end{split}
\label{eq18}
\end{equation}
which further reduces the number of truly independent components of the stress energy tensor to the set $\{\rho,P,S_i,\sigma_{ij}\}$.
Before proceeding to apply the formalism to a concrete scenario of modified gravity, it can be instructive to briefly discuss how it works in standard GR. To this aim, let us expand the to the first order in $h_{\mu\nu}$ the Riemann and Ricci curvature tensor, i.e.,
\begin{align}
R^{(1)}_{\rho\sigma\mu\nu} & =\frac{1}{2}\leri{\partial_{\sigma}\partial_{\mu}h_{\rho\nu}+\partial_{\rho}\partial_{\nu}h_{\sigma\mu}-\partial_{\sigma}\partial_{\nu}h_{\rho\mu}-\partial_{\rho}\partial_{\mu}h_{\sigma\nu}}\label{Riemann_lin}\\
R^{(1)}_{\mu\nu} & =\frac{1}{2}\leri{\partial_{\mu}\partial_{\rho}h\indices{^{\rho}_{\nu}}+\partial_{\nu}\partial_{\rho}h\indices{^{\rho}_{\mu}}-\partial_{\mu}\partial_{\nu}h-\Box h_{\mu\nu}}\label{Ricci_ten_lin},
\end{align}
where the trace $h$ is defined from the Minkowski metric, i.e., $h\equiv \eta^{\mu\nu}h_{\mu\nu}$, and $\Box\equiv \partial^\mu\partial_\mu$ denotes the flat d'Alambert operator. It follows from \eqref{Ricci_ten_lin} that the Ricci scalar reads as
\begin{equation}
R^{(1)}=\eta^{\mu\nu}R^{(1)}_{\mu\nu}=\partial_{\mu}\partial_{\nu}h^{\mu\nu}-\Box h.
\label{Ricci_sc_lin}
\end{equation}

Then, by taking advantage of the decomposition just discussed, we rewrite the Ricci tensor and the Ricci scalar in terms of gauge-invariant {quantities, namely,}
  \begin{align}
        R^{(1)}_{00}&=\bigtriangleup\Phi-\dfrac{3}{2}\ddot{\Theta} \\
        R^{(1)}_{0i}&=-\dfrac{1}{2}\bigtriangleup \Xi_i-\partial_i \dot{\Theta} \\
        R^{(1)}_{ij}&= -\partial_{(i}\dot{\Xi}_{j)}-\partial_i\partial_j \left (\Phi+\dfrac{1}{2}\Theta\right)-\dfrac{1}{2}\Box\left(\delta_{ij}\Theta+h_{ij}^{TT}\right)\\
        R^{(1)}&=3\ddot{\Theta}-2\bigtriangleup(\Theta+\Phi),
        \end{align}
which when plugged into the well-known Einstein equations $G_{\mu\nu}=\kappa T_{\mu\nu}$ and accounting for the conservation law for the stress energy tensor leads us to {the system}

\begin{align}
\bigtriangleup \Phi&=\dfrac{\kappa}{2}\left( 3P+ \rho-3\dot{S} \right)\label{phi}\\
 \bigtriangleup\Theta &=-\kappa \rho\label{theta} \\
 \bigtriangleup \Xi_i &=-2\kappa S_i \label{xi}\\
\label{htt} \Box h^{TT}_{ij}&=-2\kappa\sigma_{ij}.
\end{align}

We see that only the transverse and traceless part $h_{ij}^{TT}$ satisfies a wave equation, while the other components are just gauge degrees of freedom with no radiative behavior as they obey Poisson (Laplace in vacuum) equations. Now, when we deal with alternative theories of gravity, we expect in the linearized Einstein equations an additional contribution depending on combinations of the curvature invariants and possibly some new fields we introduced in our theoretical setting, such as scalar or vector degrees of freedom coupled in different ways to the metric. The idea behind the gauge-invariant formulation is then to seek for a suitable set of variables whose equations could be still rearranged in the form \linebreak \eqref{phi}--\eqref{htt}, provided some redefinition of the gauge-invariant components be performed. To be more specific, let us assume our theory be only endowed with an extra scalar degree along with a novel vector field, whose equations of motion can be generically described at the linearized level as
\begin{equation}
    \mathcal{E}^\varphi=0,\quad\mathcal{E}^A_\mu=0,
    \label{equation additional}
\end{equation}
where $\mathcal{E}^\varphi,\mathcal{E}^A_\mu$ enclose differential operators at most order two (we are considering in the broadest sense, the scenario of Horndeski and generalized Proca theories~\cite{Horndeski:1976gi,Tasinato:2014eka,Allys:2015sht,Allys:2016jaq,Rodriguez:2017ckc,GallegoCadavid:2019zke,Allys:2016kbq,GallegoCadavid:2020dho,GallegoCadavid:2021ljh,Rodriguez:2017wkg,Gomez:2019tbj,Garnica:2021fuu,Heisenberg:2014rta,Heisenberg:2016eld,Heisenberg:2018acv,BeltranJimenez:2013btb}), depending on the d'Alembert operator and containing eventually terms stemming from a potential in the Lagrangian. In this case, it seems reasonable to define a new set of gauge-invariant variables given by
\begin{align}
    &\Phi^{(\varphi)}\equiv\Phi+a \varphi\\
    &\Theta^{(\varphi)}\equiv\Theta+b\varphi\\
    &\Xi_i^{(A)}\equiv\Xi_i+c A_i,
\end{align}
where $a,b,c$ are coefficients related to the structure of the model, such that the {following~hold:}
\begin{align}
\bigtriangleup \Phi^{(\varphi)}&=\dfrac{\kappa}{2}\left( 3P+ \rho-3\dot{S} \right)\label{phimod}\\
 \bigtriangleup\Theta^{(\varphi)} &=-\kappa \rho\label{thetamod} \\
 \bigtriangleup \Xi_i^{(A)} &=-2\kappa S_i \label{ximod}\\
\label{httmod} \Box h^{TT}_{ij}&=-2\kappa\sigma_{ij}.
\end{align}

Thus, even if $\Phi^{(\varphi)},\Theta^{(\varphi)},\Xi_i^{(A)}$ are still nonradiative, they now actually contain dynamical degrees of freedom according to \eqref{equation additional} and this can be demonstrated to be capable of exciting additional polarizations of the gravitational wave spectrum. Let us consider, indeed, a sphere of test masses and let us analyze the effect due to the crossing of a gravitational wave, i.e., let us study the linearized geodesic deviation equation
\begin{equation}
    \frac{d^2 \delta X^{i}}{dt^2} \simeq - R\indices{^{i}_{0j0}} X_0^{j},
\end{equation}
where $\delta X^{i}$ is the displacement induced by the gravitational wave with respect to the initial position $X_0^{i}$ and we select the comoving frame where the four-velocity is normalized as $U^{\mu} = (-1, 0, 0, 0)$. Straight computation shows that the Riemann components can be locally expressed in terms of gauge-invariant quantities as

\begin{equation}
    R_{i0j0}=-\dfrac{1}{2}\ddot{h}_{ij}^{TT}+\partial_i\partial_j\Phi-\dfrac{1}{2}\delta_{ij}\ddot{\Theta}+\partial_{(i}\dot{\Xi}_{j)}.
    \label{geod}
\end{equation}

It is clear that in GR, where the only dynamical part is provided by the purely tensor modes $h_{ij}^{TT}$, we just recover the standard plus and cross polarizations. In this case, however, $\Theta$, $\Phi$, and $\Xi_i$ are not static degrees and the Poisson Equations \eqref{phimod}--\eqref{ximod} actually hold for the modified set of gauge-invariant variables $\Phi^{(\varphi)},\;\Theta^{(\varphi)}$, and $\Xi_i^{(A)}$. This implies that we can rearrange \eqref{geod} in the more convenient form
\begin{equation}
    R_{i0j0}=\underbrace{\partial_i\partial_j\Phi^{(\varphi)}-\dfrac{1}{2}\delta_{ij}\ddot{\Theta}^{(\varphi)}+\partial_{(i}\dot{\Xi}_{j)}^{(A)}}_{\text{Static part}}+\underbrace{ \leri{\frac{b}{2}\,\delta_{ij}\partial_t^2-a\,\partial_i\partial_j }\varphi-c\,\partial_{(i}\dot{A}_{j)} -\dfrac{1}{2}\ddot{h}_{ij}^{TT}}_{\text{Radiative part}},
\end{equation}
so that we expect, in general, that the evolution of $\varphi$ and $A$ could give rise to novel phenomenological effects besides the well-established tensorial strains. To be more specific, let us focus on the radiative part of the gravitational wave, which we assume to propagate in all its components along the $z$ axes and let us switch off for the sake of clarity the standard tensor modes. The geodesic deviation equation then {reads as}
\begin{align}
&\frac{d^2 \delta X}{d t^2}\simeq-\frac{b}{2}\, \partial_t^2\varphi \;X_0+\frac{c}{2}\,\partial_z\dot{A}_x\;Z_0 \label{geodesic additional pol z1}\\
&\frac{d^2 \delta Y}{d t^2}\simeq-\frac{b}{2}\, \partial_t^2\varphi \;Y_0+\frac{c}{2}\,\partial_z\dot{A}_y\;Z_
0\\
&\frac{d^2 \delta Z}{d t^2}\simeq\lerisq{\leri{-\frac{b}{2} \,\partial_t^2+a\,\partial_z^2}\varphi+c\,\partial_z\dot{A}_z}Z_0+\frac{c}{2}\leri{\partial_z\dot{A}_x\,X_0+\partial_z\dot{A}_y\,Y_0}. \label{geodesic additional pol z3}
\end{align}

We see that the scalar field $\varphi$ is now responsible for a breathing mode in the transverse plane $xy$, i.e., a conformal polarization with no preferred directions in the plane, together with a longitudinal polarization along the direction of propagation of the gravitational wave. The vector field $A_\mu$, instead, induces cross polarizations in the planes $xz$ and $yz$, commonly denoted as vector-$x$ and vector-$y$ polarizations, and an additional contribution to the longitudinal mode. We stress, however, that this very type of excitation depends crucially on the mass terms eventually present in \eqref{equation additional} (for details concerning the way of restoring the $U(1)$ gauge-invariance of $A_\mu(x)$ in the presence of a mass term via the Stueckelberg trick, see~\cite{Heisenberg:2014rta,Blas:2009yd,Ruegg:2003ps,BeltranJimenez:2016rff,Hinterbichler:2015soa}). If we assume, for instance, that \eqref{equation additional} have the simple form of Klein--Gordon, respectively, Proca equations, i.e.,
\begin{equation}
    (\Box-M_\varphi^2)\varphi=0,\quad (\Box-M_A^2)A_\mu=0,
\end{equation}
we see that in the limit $M_{A}\to 0$, the transversality condition on the vector $A_\mu$ compels us to set $A_z=0$, so that its contribution simply reduces to the vector modes. When we perform the same limit for the scalar field, instead, the perturbation along the $z$ axes cancels out only if the identity $2a=b$ holds, so that the massless condition cannot guarantee for itself the absence of a longitudinal mode. We emphasize, however, that strictly speaking, this holds even for the vector polarizations, since both the vector modes are always endowed with strains along the direction of propagation of the gravitational perturbation. We conclude, therefore, that longitudinality, in the broader sense just discussed, is quite a common feature whenever additional polarizations are taken into account. Now, for future convenience (see Section~\ref{sec3}), let us consider by way of example a generic Horndeski model with an additional scalar degree of freedom~\cite{Horndeski:1974wa,Hou:2017bqj}, described by the action \begin{equation}
    S=\frac{1}{2\kappa}\int d^4x\sqrt{-g}\sum_{i=2}^5 L_i,
\end{equation}
where the terms $L_i$ are given by
\begin{equation}
    \begin{split}
        &L_2=K(\varphi,X)\\
        &L_3=-G_3(\varphi,X) \Box\varphi\\
        &L_4=G_4(\varphi,X)R+G_{4,X}\leri{(\Box\varphi)^2-\varphi_{\mu\nu}\varphi^{\mu\nu}}\\
        &L_5=G_5(\varphi,X)G_{\mu\nu}\varphi^{\mu\nu}+\frac{1}{6} G_{5,X}\leri{(\Box\varphi)^3-3\Box\varphi\,\varphi_{\mu\nu}\varphi^{\mu\nu}+2\varphi\indices{^\mu_\nu}\varphi\indices{^\nu_\rho}\varphi\indices{^\rho_\mu}},
        \label{horndeski terms}
    \end{split}
\end{equation}
and we introduce the notation
\begin{equation}
    X\equiv-\frac{1}{2}\nabla_\mu\varphi\nabla^\mu\varphi,\quad\varphi_{\mu\nu}\equiv\nabla_\mu\nabla_\nu\varphi.
\end{equation}

Then, by expanding all the fields and taking into account perturbations at the first order, we can obtain the system of linearized equations (in vacuum):
\begin{align}
    &G_{\mu\nu}^{(1)}-\frac{G_{4,\varphi}(0)}{G_4(0)}\leri{\partial_\mu\partial_\nu-\eta_{\mu\nu}\Box}\phi=0\label{g wave equations linearized horndeski}\\
    &\leri{\Box-M^2}\phi=0,
   \label{phi wave equations linearized horndeski}
\end{align}
with the effective mass of the scalar mode given by
\begin{equation}
    M^2=-\frac{K_{,\varphi\varphi}(0)}{K_{,X}(0)-2G_{3,\varphi}(0)+\frac{3G_{4,\varphi}^2(0)}{G_4(0)}},
\end{equation}
where the functions are all evaluated in background values, $\varphi=\varphi_0,\;X=0$. We stress that the observation of the multimessenger signals associated with GW170817-GRB170817A~\cite{LIGOScientific:2017vwq,LIGOScientific:2017zic} put severe constraints on gravitational waves speed propagation in vacuum, restricting the form of the terms actually feasible of \eqref{thetamod} (see~\cite{Ezquiaga:2017ekz, Langlois:2017dyl, Creminelli:2017sry, Sakstein:2017xjx} for more details and implications in cosmological settings). With respect to~\cite{Moretti:2020kpp}, we are not considering the coupling with the matter (see the discussion in Section~\ref{sec3} or \cite{Hou:2017bqj}), since we are only interested in the phenomenology of gravitational wave polarizations on the sphere of test masses, i.e., we are neglecting at all the backreaction due to the displacement of the medium particles. It is important to note that the form of \eqref{g wave equations linearized horndeski} and \eqref{phi wave equations linearized horndeski} encompass a large variety of modified theories of gravity, such as metric $f(R)$ models~\cite{Liang2017,Moretti2019} or generalized hybrid metric-Palatini gravity~\cite{Bombacigno:2019did,Rosa:2021lhc}, even if in this very last case, we deal with two additional scalar fields whose interaction can lead to interesting phenomena such as beatings (see~\cite{Bombacigno:2019did} for more details). 
Now, a bit of manipulation reveals that the coefficients $a,\,b$ take the simple~form
\begin{equation}
    2a=b=\frac{G_{4,\varphi}(0)}{G_4(0)},
\end{equation}
so that \eqref{geodesic additional pol z1}--\eqref{geodesic additional pol z3} {simplifies to}
\begin{align}
&\frac{d^2 \delta X}{d t^2}\simeq-\frac{b}{2}\, \partial_t^2\varphi \;X_0 \label{geodesic scalar general1}\\
&\frac{d^2 \delta Y}{d t^2}\simeq-\frac{b}{2}\, \partial_t^2\varphi \;Y_0\\
&\frac{d^2 \delta Z}{d t^2}\simeq-\frac{b}{2}\leri{ \partial_t^2-\partial_z^2}\varphi\;Z_0, \label{geodesic scalar general3}
\end{align}
which, under the hypothesis of a monochromatic wave perturbation with planar wave front, described by the dispersion relation $\omega^2=k^2+M^2$, can be further {rearranged as}

\begin{align}
&\frac{d^2 \delta X}{d t^2}\simeq \frac{b \omega^2}{2}\, \varphi \;X_0\\
&\frac{d^2 \delta Y}{d t^2}\simeq \frac{b \omega^2}{2}\, \varphi \;Y_0\\
&\frac{d^2 \delta Z}{d t^2}\simeq\frac{bM^2}{2}\,\varphi\;Z_0.
\end{align}

In this form, some peculiar characteristic are easily displayed. Firstly, we note as in the limit $M\to 0$, the transversality is correctly recovered, while the requirement of having no additional perturbation at all implies $b=0$, which when reformulated in terms of the function $G_4(\varphi,X)$ leads us to consider only minimal coupling of the scalar field with the Ricci scalar. Secondly, we observe for $b\neq 0$ that the two stresses are always in phase, with the amplitude of the breathing mode greater than the amplitude of the longitudinal one.

\section{Gravitational Waves in Matter}\label{sec3}
In this section, we address the broad and debated issue of the propagation of gravitational waves within matter. The problem has a long history in the literature and the great number of works on the theme can be divided into two main categories, depending on the approach to the characterization of the material medium. In fact, the latter can be described either by continuous fields or by a large number of particles, whose statistics are depicted by a distribution function. In the first case, we deal with a hydrodynamic approach, and the dynamics can be traced, in principle, with fully analytic methods, whilst in the second, the problem is treated by following kinetic theory prescriptions, resulting in an intrinsically statistic picture. Let us enumerate the most important results achieved by following the hydrodynamic approximation. In~\cite{Hawking:1966qi,Madore:1972ww,Madore,Prasanna:1999pn}, the amplitude damping of gravitational perturbations either on an expanding or generic background was analyzed. The main result is that damping occurs only when the propagation in a dissipative fluid, characterized by a definite viscosity, is considered. The same issue is addressed also 
in~\cite{Anile}, where the thorny issue of the gauge freedom pertaining to the wave throughout the propagation in the medium, strictly connected to the count of the physical and fictitious degrees of freedom, is carefully examined. The authors demonstrate that there exists two physical polarization states, as in vacuum, and confirm again the previous results regarding the characteristic time of absorption. An interesting phenomenological application following from these findings is given in~\cite{Goswami:2016tsu}, where the authors illustrate that observations of gravitational waves signals through LIGO--Virgo interferometers can constrain the viscosity of the cosmological fluid. In~\cite{Barta:2017vip}, a modified dispersion relation and extra modes of polarizations are shown to appear when the medium considered is a spherical cloud (perfect fluid in Schwarzschild metric); the case of pressureless matter (dust) is studied in~\cite{Ehlers:1987nm} through the application of a WKB approximation. The authors derive two different dispersion relations: the first is shown to be simply degenerated and corresponding to tensorial perturbations (gravitational waves), whilst the second is characterized by a double degeneracy, describing the dispersion of both density and vorticity perturbations, i.e., novel degrees of freedom arising when the propagation of perturbations in matter is considered.

Further, in the case of kinetic approach to the description of the material medium, a brief review of the main results achieved can be provided. For instance, in~\cite{osti_4641583,Chesters:1973wan} is considered the interaction of transverse traceless gravitational waves with noncollisional particles on a Minkowski background. In both works, it is recognized that Landau damping~\cite{Landau:1946jc} arises only if the wave phase velocity is subluminal throughout the propagation within the medium. Moreover, it is shown that such a condition is fulfilled only by including anisotropies in the distribution function describing the equilibrium configuration. The case of collisionless particles on a globally flat FLRW background is considered in~\cite{PhysRevD.13.2724,Gayer:1979ff,1978SvA....22..528Z}. In all three works, it is assumed that the equilibrium configuration of the medium is described by a J\"uttner distribution~\cite{1911AnP...339..856J} and the appearance of extra modes of oscillations, corresponding to scalar and vector perturbations, is highlighted. The results in~\cite{PhysRevD.13.2724,Gayer:1979ff}, derived under the assumption of frequency of the wave much larger than the Hubble parameter, show that no damping is expected. On the contrary, the analysis in~\cite{1978SvA....22..528Z}, pursued without the aforementioned assumption, proves that vector perturbations were strongly suppressed in the early Universe (the damping rate is found to be proportional to $\eta^{-4}$, being $\eta$ the comoving time). The role of collisions is examined in~\cite{Ignatyev:2010zu,Baym:2017xvh}. Particularly, in~\cite{Ignatyev:2010zu} is quantified the damping of gravitational waves coming from scattering processes between elementary particles. In~\cite{Baym:2017xvh} is described the arising of a peculiar mechanism: on one hand, collisions macroscopically act as viscosity, causing the appearance of damping, but on the other hand, they tend to suppress anisotropies, which---as already mentioned---are a source of damping. Therefore, the late-time evolution is substantially undamped and only primordial perturbations are significantly suppressed. The case of the interaction of cosmological gravitational waves with neutrinos is considered in~\cite{Weinberg:2003ur,Lattanzi:2005xb,Lattanzi:2010gn,Benini:2010zz}. Specifically, in~\cite{Weinberg:2003ur,Benini:2010zz}, neutrinos are assumed to be already decoupled from the cosmological bath; therefore, collisions are ignored. The maximum damping is calculated for perturbations that enter the horizon during the radiation-dominated era and equals roughly $35 \%$, independently from cosmological parameters and frequencies of the radiation. Collisions are included in~\cite{Lattanzi:2005xb} in order to describe the interaction of primordial gravitational waves with neutrinos that are still in equilibrium with the cosmological bath. It is found that the amount of absorbed radiation is proportional to the abundance of neutrinos and is independent from the frequency considered or the details of the collisions. The transition from collisional to collisionless regimes is addressed in~\cite{Lattanzi:2010gn}, where it is shown that, during the transition, the damping is frequency-dependent. In particular, the appearance of a characteristic spectral signature in the region of the $\text{nHz}$ is outlined. An analytic treatment of the integro-differential equation obtained in~\cite{Weinberg:2003ur} is provided in~\cite{PhysRevD.88.083536}, where asymptotic solutions are evaluated in terms of Bessel functions. In~\cite{Milillo:2008qu}, a term that gives account for the spin of the particles is included in Vlasov equation. However, explicit calculations show that the spin contribution does not couple with traceless transverse tensorial metric perturbations. Lastly, in~\cite{Flauger:2017ged} is considered the case of the interaction of gravitational waves with cold dark matter. It is shown that the dominant effect at astrophysical scale is constituted by a frequency-dependent modification in the propagation speed of the wave. Even though it is outlined that, in principle, the power spectrum of primordial gravitational waves carries signatures of the interaction with dark matter, the effects are found to be small, due to the highly nonrelativistic nature of the medium traversed.

A third and far less-pursued approach to the description of the material medium is the so-called molecular picture. In this case, the matter content is imagined as a set of pointlike particles that can be grouped into molecules. The external microscopic field, i.e., the gravitational wave, alters the structure of the molecule, generating an induced quadrupole moment. At a macroscopic level, the average quadrupole moment of the whole medium is then responsible for peculiar modifications in the gravitational wave dynamics, as for instance dispersion or damping. This line of research begins with the work from Szekeres~\cite{Szekeres:1971ss}, in which is provided the definition of the quadrupole polarization tensor of the molecule in terms of the separation vectors between point-particles and centers of mass of the molecules. In addition to this, it is found that when the Weyl tensor is assumed to be the fundamental dynamical variable, dispersion occurs and no extra polarizations are excited by the propagation within matter. A prosecution of this work is provided in~\cite{Montani:2003vm}, with a particular focus on the static limit, i.e., the modified Newtonian potential affected by the contribution of the average quadrupole moment of the medium. Lastly, in~\cite{Svitek:2008pd} is outlined the possibility of the settling of an amplitude damping, when the molecules are treated on an anti-de-Sitter background.
In the next sections, we will illustrate two works regarding the propagation of gravitational waves in matter. In the first~\cite{Montani:2018iqd}, it is shown that gravitational waves from general relativity are endowed with five physical degrees of freedom when propagating in a medium of massive spherical molecules. The analysis is pursued in the theoretical framework of macroscopic gravity, and the constitutive relation between the molecule quadrupole tensor and the external gravitational wave is derived from a simple phenomenological model. The number of radiating degrees of freedom and the dispersion suffered by the macroscopic radiation allow us to establish an effective analogy with the propagation of massive gravitational waves, as described by the Fierz--Pauli theory~\cite{Fierz:1939ix}---the linear approximation of the full theory of massive gravity, which has been introduced only in recent years~\cite{deRham:2010kj}. In the second work~\cite{Moretti:2020kpp}, the authors analyzed the interaction between gravitational waves from Horndeski theories and a noncollisional medium of massive particles. The kinetic treatment enforced outlines the presence of Landau damping for the scalar massive mode, when an inequality between theory parameters and physical quantities describing the medium is satisfied.   
\subsection{Macroscopic Gravity}
Longitudinal stresses induced by gravitational waves are not a specific trait of alternative theories of gravity, but can arise also when the propagation of gravitational waves from general relativity within matter is addressed. Indeed, it is a well-known fact that the TT-gauge, which manifests the transversality of the two physical degrees of freedom $h_+$ and $h_\times$, is achievable only in vacuum. When the general case is considered, instead, the transverse nature of gravitational waves is not ensured and the number of physical degrees of freedom must be deduced from the equation of motion. With respect to the discussion in Section~\ref{sec2}, where it was shown that in the matter, we have to deal with spurious gauge degrees of freedom as well; here, we refer to the description of the interaction with the traversed medium in a self-consistent way. In other words, we are interested in expressing the perturbation in the stress-energy tensor due to the presence of the wave through a constitutive relation, i.e., in terms of the components of the wave itself. The resulting equation of motion is closed in the metric perturbation and, in general, exhibits a reduced gauge freedom with respect to the vacuum case, which can lead to fundamental differences in the number of physical degrees of freedom. In this case, the typical Poisson equations ruling the scalar and vector components of the metric perturbation are drastically affected, resulting in novel radiative modes besides the tensor polarizations. We shall consider, in particular, the case of gravitational waves from general relativity traveling in a material medium of pointlike particles grouped into molecules. In~\cite{Szekeres:1971ss}, it is demonstrated that, under covariant averaging, it is possible to extract the molecular moments from the general matter distribution. The first-order expression of the macroscopic field equation is~then 
\begin{equation}\label{macrofe}
    \left < G_{\mu\nu} \right >= \kappa \left (T_{\mu\nu}^{(f)}+\frac{1}{2}Q_{\mu\rho\nu\sigma , }^{\qquad \rho \sigma} \right) ,
\end{equation}
where $\left < \cdot \right >$ is the covariant averaging applied to the microscopic field equation, $T_{\mu\nu}^{(f)}$ is the free stress-energy tensor of the pressureless dust composed by the molecules' centers of mass and $Q_{\mu\rho\nu\sigma}$ is the quadrupole tensor, describing the molecule structure. An expression of the latter is also provided in~\cite{Szekeres:1971ss}, given in terms of the separation vectors between each pointlike particle and the center of mass of the relative molecule. The external (microscopic) gravitational wave is responsible for a perturbation in the trajectories of the pointlike particles, which can be calculated in terms of the amplitude of the wave itself from the geodesic deviation equation. When the relation is inserted into the expression of the quadrupole tensor, the set of constitutive relations 
\vspace{-12pt}
\begin{align}
    Q_{i0j0}&=\frac{1}{2}\epsilon_g h_{ij,00} \label{matrel1}\\
    Q_{0ijk}&=0 \label{matrel2}\\
    Q_{ijkl}&=0 \label{matrel3}
\end{align}
is obtained. The parameter $\epsilon_g$ is the gravitational dielectric constant, quantifying the amount of induced quadrupole moment with respect to the strength of the external wave. Its value is determined by physical parameters characterizing the molecule
\begin{equation}
    \epsilon_g=\dfrac{MNL^2}{4 \omega_0^2},
\end{equation}
with $M$ and $L$ are the molecule mass and radius, respectively; $N$ is the number of molecules per unit volume; and $\omega_0$ is the proper frequency of the molecule, defined in terms of the molecule mass density $\rho$ as 
\begin{equation}\label{omegapropria}
   \omega_0 =\sqrt{\dfrac{\kappa \rho}{6}}.
\end{equation} 

Let us describe in detail the form of the macroscopic field equation descending from the adoption of the set of constitutive relations \eqref{matrel1}--\eqref{matrel3}. The first aspect that has to be remarked is that the macroscopic field Equation \eqref{macrofe} is endowed with a global diffeomorphism invariance, due to the covariant nature of the quadrupole tensor. Therefore, it is possible to require that the averaged metric perturbation $\left<\bar{h}_{\mu\nu}\right >$---which is nothing more than the usual trace-reversed matrix $\left<\bar{h}_{\mu\nu}\right >=\left<h_{\mu\nu}\right >-\frac{1}{2}\eta_{\mu\nu}\left<h\right >$---must satisfy the Hilbert gauge condition
\begin{equation}
    \partial^\mu \left<\bar{h}_{\mu\nu}\right >=0.
\end{equation}

As a result, the ten free components pertaining to the general rank two symmetric tensor $\left<\bar{h}_{\mu\nu}\right >$ are reduced to six. In particular, by choosing the reference frame in order to have the direction of propagation of the perturbation along the $z$ axis and writing the components of the perturbation in the Fourier space, the following matrix is obtained
	\begin{equation}
\bar{h}_{\mu\nu}=\left(\begin{matrix}\left( \frac{k}{\omega}\right)^2\bar{h}_{33}& -\frac{k}{\omega}\bar{h}_{13}  & -\frac{k}{\omega}\bar{h}_{23} & -\frac{k}{\omega}\bar{h}_{33} \\\vdots&\bar{h}_{11} & \bar{h}_{12}&\bar{h}_{13} \\ \vdots & \cdots & \bar{h}_{22}&\bar{h}_{23} \\ \vdots &\cdots &\cdots & \bar{h}_{33}\end{matrix}\right),
\end{equation}
where the symbol $\left < \cdot \right >$ is dropped, here and in the following, for the sake of convenience. An ulterior gauge transformation that could further reduce the number of free components of the metric perturbation is not, in general, achievable. Indeed, it is a well-known result that, once Hilbert gauge fixing has been enforced, one can cancel out a certain number of components if, and only if, they result to be a solution of d'Alembert equation. Therefore, the analysis of the equations of motion is 
the way to reveal the number of physical degrees of freedom of the theory. Let us display the $(00)$, $(0i)$, and $(ij)$ components of the field equation, which read, respectively, 
\begin{equation}
\begin{split}
  \Box \bar{h}_{33} &= -4 \pi G \epsilon_g\, \partial_0^4 \,\bar{h}_{33}+2\pi G \epsilon_g\dfrac{\omega^2}{k^2}\bigtriangleup\partial_0^2\,\bar{h} \\
  \Box\bar{h}_{3i} &=-4\pi G \epsilon_g\,\partial_0^4\,\bar{h}_{3i}-2\pi G \epsilon_g\dfrac{\omega}{k}\,\partial_0^3 \partial_i \bar{h} \\
  \Box \bar{h}_{ij} &=-4\pi G \epsilon_g\,\partial_0^4\,\bar{h}_{ij}+2\pi G \epsilon_g\eta_{ij} \partial_0^4\, \bar{h}.
\end{split}
\end{equation}

From these, it is possible to calculate a wave equation for the trace $\bar{h}$, resulting in
\begin{equation}
\Box\bar{h}=-2\pi G\epsilon_g\left(\dfrac{k^2}{\omega^2}-1\right)\partial_0^4\,\bar{h}.
\end{equation}

The descending dispersion relations $\omega=\omega(k)$ are easily calculated 
\begin{equation}
\omega(k)= \pm k \quad \quad  \omega(k)=\pm \sqrt{2}m,
\end{equation}
where it has been introduced the constant wavenumber $m=\leri{4\pi G\epsilon_g}^{-\frac{1}{2}}$. This finding shows that the trace can be written as the superposition of a solution of d'Alembert equation, which, as noted before, can be canceled out with a gauge transformation that preserves Hilbert gauge, and a local oscillation, devoid of propagative features (in other words, the group velocity $\frac{d\omega}{dk}$ is strictly null for the second branch). Hence, we are free to set $\bar{h}=0$, bringing the count of degrees of freedom to five. After this fixing, the remaining free components are solutions of the equation
\begin{equation}\label{waveeqmacro}
	\Box\bar{h}_{\mu\nu}=-\dfrac{1}{m^2}\partial_0^4\,\bar{h}_{\mu\nu}.
\end{equation}

It is immediate to notice that the fourth-order wave equation obtained for the five macroscopic degrees of freedom is not manifestly covariant: the reason for this fact is that the constitutive relations \eqref{matrel1}--\eqref{matrel3} have been calculated in a definite reference frame, i.e., the one that is comoving with the center of mass of the molecule. As is well-known, any higher-derivative theory can be affected by the Ostrogradsky instability, resulting in an energy spectrum unbounded from below. A rigorous proof of the global stability of the macroscopic theory should be carried out by showing the singularity of the connected effective Lagrangian; however, in our case, the latter cannot be defined, due to the lack of covariance of the equation of motion. Therefore, we limit our analysis to a phenomenological level and we calculate the dispersion relation coming from \eqref{waveeqmacro}, \linebreak which~reads 
\begin{equation}\label{reldispmacro}
\omega^2_{\pm}(k)= -\dfrac{m^2}{2}  \pm \sqrt{\dfrac{m^4}{4}+m^2k^2}.
\end{equation}

As it can be easily inferred, the plus-signed branch of the dispersion relation implies pure dispersion of the signal, whereas the minus-signed is related to damping and enhancement phenomena. However, by performing the limit in which the material medium is removed $\epsilon_g \to 0$ (which is equivalent to $m \to \infty$), one finds that only $\omega_{+}(k)$ goes back to be a vacuum dispersion relation $\omega(k)=\pm k$. On the contrary, the minus-signed branch has a divergent behavior in the same limit: this suspiciously nonphysical feature allows us to consider the solutions characterized by $\omega_{-}(k)$ as fictitious. Hence, in the following, we will focus only on purely dispersive solutions. We calculate the following group velocity
\begin{equation}
v_g(k)=\dfrac{m^2 k}{2\sqrt{\left(\sqrt{m^2k^2+\dfrac{m^4}{4}}-\dfrac{m^2}{2}\right)\left(m^2k^2+\dfrac{m^4}{4}\right)}},
\end{equation}
which, in the long-wavelengths limit $k \ll m$, reduces to the approximate form 
\begin{equation}\label{vgapprox}
v_g(k) \simeq1-\dfrac{3k^2}{2m^2}.
\end{equation} 

We outline that the group velocity shows the good behavior of being subluminal for all wavenumbers. We are now interested in describing the polarization content of the five macroscopic degrees of freedom: we proceed to exploit the condition $\bar{h}=0$ via gauge~fixing
\begin{equation}
 \begin{split}
 \bar{h}_{11}&=\bar{h}_++\bar{h}^*\\
 \bar{h}_{22}&=-\bar{h}_++\bar{h}^*\\
\bar{h}_{33}&=\dfrac{2\omega^2}{k^2-\omega^2}\bar{h}^*.
\end{split}
 \end{equation} 
 
 As seen in the previous section, we study the linearized geodesic deviation equation in the comoving frame
 
 \begin{equation}
    \frac{d^2 \delta X^{i}}{dt^2} = - R\indices{^{i}_{0j0}} X_0^{j}
\end{equation}
 separately for each free component of $\bar{h}_{\mu\nu}$. First, we recover the standard plus and cross tensorial polarizations carried by the components named $\bar{h}_+$ and $\bar{h}_{12}$. Then, we display the sets of equations of motion, which are obtained when the components $\bar{h}_{13}$ and $\bar{h}_{23}$ are kept non-null:   
\begin{equation}
 	\begin{split}
 	\dfrac{d^2 \delta X}{dt^2}&=\dfrac{k^2-\omega^2}{2}\bar{h}_{13}Z_0\\
 	\dfrac{d^2 \delta Y}{dt^2}&=0 \\
 	\dfrac{d^2 \delta Z}{dt^2}&=\dfrac{k^2-\omega^2}{2}\bar{h}_{13}X_0
 	\end{split}
 	\end{equation}
 	
 	\begin{equation}
 	\begin{split}
 	\dfrac{d^2 \delta X}{dt^2}&=0\\
 	\dfrac{d^2 \delta Y}{dt^2}&=\dfrac{k^2-\omega^2}{2}\bar{h}_{23}Z_0\\
 	\dfrac{d^2 \delta Z}{dt^2}&=\dfrac{k^2-\omega^2}{2}\bar{h}_{23}Y_0.
 	\end{split}
 	\end{equation}
 	
 	We recognize that the displacements induced by these components on a sphere of test particles are equivalent to vector-$x$ and vector-$y$ polarizations, respectively (see \mbox{Section~\ref{sec2}}). Eventually, when we plug the component $\bar{h}^*$ into the geodesic deviation equation, we obtain the following
 	\begin{equation}
 	\begin{split}
 	\dfrac{d^2 \delta X}{dt^2}&=-\dfrac{\omega^2}{2}  \bar{h}^* X_0\\
 	\dfrac{d^2 \delta Y}{dt^2}&=-\dfrac{\omega^2}{2}  \bar{h}^* Y_0  \\
 	\dfrac{d^2 \delta Z}{dt^2}&=-\left(k^2-\omega^2\right)  \bar{h}^* Z_0.
 	\end{split}
 	\end{equation}
 	
 As it can be deduced from a direct confrontation with \eqref{geodesic scalar general1}--\eqref{geodesic scalar general3}, the component $\bar{h}^*$ is responsible for the simultaneous excitation of a longitudinal polarization superposed to a breathing in the transverse plane. Again, by taking into account the dispersion 
 relation~\eqref{reldispmacro}, it follows that the two stresses are always in phase, with amplitude of the breathing greater than the amplitude of the longitudinal.  

The results collected in this section suggest that it is then possible to build a bridge between gravitational and electromagnetic radiation propagating in matter. Indeed, the five physical degrees of freedom of the macroscopic wave are related to the viable set of states pertaining to a massive spin 2 boson, so it is legitimate to define the concept of effective mass of the graviton, fully analogous with the effective mass acquired by photons when crossing a plasma~\cite{PhysRevE.49.3520,Mendonca:2000tk}. In the following, we will further exploit this comparison, by introducing the concepts of Landau damping for scalar gravitational waves and neutralizing background for gravitational plasmas.
\subsection{Landau Damping for Gravitational Scalar Waves}\label{sec3.2}
As previously stated, there is a great number of works in the literature that have investigated the possibility of Landau damping for tensorial gravitational waves, propagating either on a static flat background or over an expanding one. To summarize, it has been found that Landau damping can occur only by including anisotropies in the equilibrium configuration of the medium or by considering a non-null collision term in the Boltzmann equation governing the distribution function time evolution. A second important result achieved by many works on the theme is that a subluminal phase velocity for the signal within the medium results to be a necessary condition for the emergence of the damping phenomenon. With great generality, it has been demonstrated that tensorial massless perturbations on Minkowski cannot satisfy such a request. Is it now important to clarify the writers' opinion on a technical and conceptual controversy contained in a number of works on the theme: it has been argued that results obtained under the assumption of a flat Minkowski background are not able to satisfactorily describe realistic settings for the fact that, in principle, one should obtain the background metric as the solution of the unperturbed Vlasov problem, i.e., it should be determined by the equilibrium configuration of the medium. Our point of view regarding this issue is based on an analogy with plasma physics. In this field, it is recognized that Landau damping and other collective phenomena are due to the presence of a long-range interaction affecting the great number of electrons contained within a Debye length. This latter concept, in turn, emerges from the screening of the Coulomb electrons' potential caused by the ions' distribution. It is therefore evident that, in order to translate the physics of Landau damping in the gravitational sector, it must be provided a mechanism that could mimic the presence of a particle species with negative mass. We argue that it is precisely the possibility of setting a local inertial frame, covering the proximity of a chosen space-time event, which gives the aforementioned phenomenological resemblance between electromagnetic plasma physics and its gravitational counterpart. Indeed, in gravity, one can always make the Christoffel symbols relative to the background almost vanish in a certain neighborhood of a space-time-point, canceling out---for all the particles contained in that region---the mean field generated by the whole medium. We further point out how it is only in such conditions that the assumption of isotropy of the equilibrium configuration of the particles, which is common to many works on the theme, is sufficiently well-grounded, given that any non-null mean field would introduce a privileged direction and inhomogeneities. Hence, in order to study the possibility of Landau damping for gravitational waves, the assumption of a Minkowski background is in all parts acceptable, as long as the considered wavelength of the radiation is smaller than the size of the space-time region to which the treatment can be enforced. Therefore, a sufficient condition for the applicability of this conceptual structure is to look at gravitational perturbations much smaller than the characteristic background scale. As mentioned above, the other fundamental ingredient for the settling of the damping phenomenon is the subluminal phase velocity for modes within the medium. As it emerges from the literature and the previous sections of this work, a subluminal speed of propagation is often related to some sort of massive behavior of the gravitational ripples---be it effective, i.e., emerging at a macroscopic level from the interaction with the medium, or intrinsic, i.e., caused by the peculiar structure of the theory. Hence, it is intriguing to look at the great number of theories of modified gravity that admit the presence of massive degrees of freedom, in order to find out if Landau damping constitutes a viable channel for energy exchange between gravitational radiation and matter.

\clearpage
In this section, we present the analysis of the propagation, in a medium composed by massive particles, of gravitational waves for Horndeski theories, under the assumption of negligible collisions. The general setting of Horndeski theories together with the form of the linearized field equations have been provided in Section~\ref{sec2}. Here, we rewrite them in the presence of a nonvanishing stress energy tensor, namely,
\begin{align}
    &G_{\mu\nu}^{(1)}-\frac{G_{4,\varphi}(0)}{G_4(0)}\leri{\partial_\mu\partial_\nu-\eta_{\mu\nu}\Box}\phi=\kappa''T_{\mu\nu}^{(1)} \label{eqcampohorn}\\
    &\leri{\Box-M^2}\phi=\kappa'T^{(1)}, \label{eqcamposcalare}
\end{align}
where the source terms $T_{\mu\nu}^{(1)}$ and $T^{(1)}$ are meant to be perturbations of order $\mathcal{O}(h)$ induced by the wave itself, whereas the coupling constants $\kappa '$ and $\kappa ''$ are expressed in terms of the theory parameters as

\begin{align}
    &\kappa'=\frac{G_{4}(0)\kappa}{G_{4}(0)\leri{K_{,X}(0)-2G_{3,\varphi}(0)}+3G_{4,\varphi}^2(0)}\\
    &\kappa''=\frac{\kappa}{G_{4}(0)}.
\end{align}

In order to separate the scalar degree of freedom from the others, it is useful to introduced the generalized trace-reversed metric perturbation
\begin{equation}
    \bar{h}_{\mu\nu}\equiv h_{\mu\nu}-\frac{1}{2}\eta_{\mu\nu}\leri{h+2 b \phi},
\end{equation}
where $b$ is the coefficient introduced in Section~\ref{sec2}.

In this manner, it is possible to recast \eqref{eqcampohorn} in the familiar form \eqref{equation gravitational wave}, and we can impose on the novel metric variable $\bar{h}_{\mu\nu}$ the Lorentz condition, i.e., $\partial_\mu\bar{h}^{\mu\nu}=0$. In the following, we will focus on the purely tensor part of $\bar{h}_{\mu\nu}$, and we will neglect its vector component. The scalar part is instead now encoded in the field $\phi$, which, as we will show, naturally decouples from the tensor perturbations.

The material medium in which gravitational waves are propagating is described by the distribution function $f(\vec{x},\vec{p},t)$, whose time evolution is governed by the Vlasov equation
\begin{equation}
  \dfrac{D f}{dt}=\dfrac{\partial f}{\partial t}+\dfrac{d x^i}{dt}\dfrac{\partial f}{\partial x^i}+\dfrac{dp_i}{dt}\dfrac{\partial f}{\partial p_i}=0
\end{equation}
and in terms of which the stress-energy tensor is defined
\begin{equation}
T_{\mu\nu}^{(1)}=\dfrac{1}{\sqrt{-g}}\int d^3p\, \dfrac{p_\mu p_\nu}{p^0}\, f (\vec{x},\vec{p},t).
\end{equation}

It must be stressed that the distribution function is assumed to be dependent from the covariant components of the momentum $p_i$ and that the volume element in the space of momenta is $d^3p=dp_1 dp_2 dp_3$. The external force $\frac{dp_i}{dt}$, driving the particles out of the equilibrium configuration, is calculated from the geodesic equation 
\begin{equation}
    \dfrac{dp^\mu}{dt}+\Gamma^\mu_{\alpha \beta} \dfrac{p^\alpha p^\beta}{p^0}=0,
\end{equation}
where $\Gamma^\mu_{\alpha \beta}$ are the Christoffel symbols, expressed up to first order in the metric perturbation; $p^0\equiv \sqrt{m^2+g^{ij}p_ip_j}$ is the particle energy; and $m$ is the particle mass. It results in 
\vspace{-8pt}
\begin{equation}
    \dfrac{dp_i}{dt}=\dfrac{1}{2p^0}\leri{p_kp_l\dfrac{\partial\bar{h}_{kl}}{\partial x^i}+b m^2 \dfrac{\partial \phi}{\partial x^i}}.
\end{equation}

We assume that gravitational waves start to interact with the matter medium at the fiducial time $t=0$, so that for any negative time, the distribution function is an isotropic solution of the unperturbed equation $f_0(p)$, where $p$ is the Euclidean modulus of the particle momentum, i.e., $p\equiv\sqrt{\delta^{ij}p_ip_j}$. Then, in order to assure the continuity of the distribution function, we impose that at $t=0$, it is simply given by $f(\vec{x},\vec{p},0)=f_0\leri{ \sqrt{g^{ij}(\vec{x},0)p_ip_j}}$. 

At first order in the metric perturbation, instead, we have 
\begin{equation}
   f(\vec{x},\vec{p},0)=f_0 \leri{p}-\dfrac{f_0'(p)}{2}\leri{\dfrac{p_ip_j}{p}\bar{h}_{ij}(\vec{x},0)-b\, p\, \phi(\vec{x},0)},
\end{equation}
where $f_0'(p) \equiv \frac{d f_0}{dp}$. It is then recognizable at this stage a fundamental discrepancy with respect to the electromagnetic analogue: at the initial time, the distribution function acquires a non-null contribution from the discontinuous wave-front, which perturbs the material medium. Then, at any positive time, the presence of the gravitational wave causes an alteration in the particles distribution, which we represent as $\delta f (\vec{x},\vec{p},t)$. The linear treatment of the Landau damping phenomenon requires that the order of magnitude of the perturbation of the distribution function with respect to the background values coincides with that of the gravitational wave, i.e., $\frac{\delta f}{f_0}=\mathcal{O}(h)$. Hence, neglecting all quadratic terms, the linearized Vlasov equation for $\delta f (\vec{x},\vec{p},t)$ is obtained 
\begin{equation} \label{linearvlasov}
    \frac{\partial \delta f }{\partial t}+\frac{p^m}{p^0}\frac{\partial \delta f}{\partial x^m}-\frac{f_0'(p) }{2p} \leri{p_ip_j\frac{\partial \bar{h}_{ij}}{\partial t}-\alpha p^2 \frac{\partial \phi}{\partial t}-b p^0 p^m \frac{\partial \phi}{\partial x^m}}=0.
\end{equation}

The system composed by \eqref{equation gravitational wave} (restricted to the solely spatial indices), \eqref{eqcamposcalare} and \eqref{linearvlasov} results are closed once the source terms for the metric perturbations are expressed in terms of $\delta f (\vec{x},\vec{p},t)$, i.e.,
\begin{align}
& T^{(1)}=-m^2 \int d^3p\, \dfrac{\delta f (\vec{x},\vec{p},t)}{p^0}\\
\label{traccia}
& T_{ij}^{(1)}=\int d^3p\, \dfrac{p_ip_j}{p^0}\delta f (\vec{x},\vec{p},t).
\end{align}

Now, we search for plane wave solutions of this differential problem: particularly, we set the $z$ axis of our reference frame to be coincident with the direction of propagation of the perturbations, so that the field's dependence from spatial coordinates is restricted to the sole $z$. Then, as it is usually done in electromagnetic plasma theory, we perform a Fourier transform of parameter $k$ on the spatial coordinate, together with a Laplace transform of parameter $s$ on the time coordinate $t$. In the Fourier--Laplace space, the differential problem is transformed into an algebraic one, so that solutions for the distribution function~perturbation
\begin{equation}
    \delta f^{(k,s)}(\vec{p})=\frac{\frac{f_0'(p)}{2 p}\leri{p_ip_j\leri{s\,\bar{h}_{ij}^{(k,s)}-\bar{h}_{ij}^{(k)}(0)}-b\leri{p^2s+ikp_3p^0}\phi^{(k,s)}+b p^2\phi^{(k)}(0)}}{s+ik \frac{p_3}{p^0}}
\label{deltafFL}
\end{equation}
and for the metric fields
\begin{align}
    &\phi^{(k,s)}=\frac{s\phi^{(k)}(0)+m^2\kappa'\int d^3p \frac{\delta f^{(k,s)} (\vec{p})}{p^0}}{s^2+k^2+M^2}
    \label{FLkleingordon}\\
    &\bar{h}_{ij}^{(k,s)}=\frac{s\bar{h}_{ij}^{(k)}(0)+2\kappa''\int d^3p \frac{p_ip_j}{p^0}\delta f^{(k,s)} (\vec{p})}{s^2+k^2}
    \label{FLtensorkleingordon}
\end{align}
are readily obtained. The Fourier and Fourier--Laplace projections are denoted by $\phi^{(k)}(t)$ and $\phi^{(k,s)}$, respectively, and analogously for the other fields. Moreover, without loss of generality, we have set $\dot{\phi}^{(k)}(0)$, $\dot{\bar{h}}_{ij}^{(k)}(0)$, and $\delta f^{(k)}(0)$ vanishing. Once the expression for the distribution function perturbation \eqref{deltafFL} is inserted into \eqref{FLkleingordon} and \eqref{FLtensorkleingordon}, it can be shown that scalar and tensor perturbations within the medium are naturally decoupled, i.e.,

\begin{align}
    &\phi^{(k,\omega)}=\dfrac{\leri{- i \omega + i b\pi m^2 \kappa'\int d\rho dp_3\, \rho\, \dfrac{f_0'(p) p }{p^0\omega-kp_3}}\phi^{(k)}(0)}{(k^2+M^2-\omega^2)\epsilon^{(\phi)}(k,\omega)}\label{equationphiuncoupled}\\
    &\bar{h}_{ij}^{(k,\omega)}=\dfrac{\leri{-i \omega-\frac{i \pi \kappa''}{2}\int d\rho dp_3 \,\rho^5 \dfrac{f'_0(p)}{p(p^0\omega-kp_3)}}\bar{h}_{ij}^{(k)}(0)}{(k^2-\omega^2)\epsilon^{(h)}(k,\omega)},
    \label{equationhuncoupled}
\end{align}
where the integrals have been converted into cylindrical coordinates by means of the substitution $\leri{p_1,p_2}\rightarrow \leri{\rho,\theta}$, with $\rho^2=p_1^2+p_2^2$ and $\tan(\theta)=\frac{p_2}{p_1}$, and we have redefined $s\equiv -i\omega$. We have also introduced the complex dielectric functions 
\begin{equation}
    \label{epsilonscalar}
\epsilon^{(\phi)}(k,\omega)=1
    +\dfrac{b \pi m^2 \kappa'}{k^2+M^2-\omega^2}\int d\rho dp_3\,\rho \dfrac{f_0'(p)}{p} \dfrac{p^2\omega-kp^0p_3}{p^0\omega-kp_3}
\end{equation}
\begin{equation}
    \label{epsilontensor}
\epsilon^{(h)}(k,\omega)=1
    -\dfrac{\pi\kappa''}{2(k^2-\omega^2)}\int d\rho dp_3\,\rho^5 \dfrac{f_0'(p)}{p} \dfrac{\omega}{p^0\omega-kp_3},
\end{equation}
whose roots are connected with the allowed oscillation modes within the medium. Indeed, as is well-known from the Laplace transform theory, the inverse Laplace transform is dominated, for long times, by the contribution stemming from the poles in the complex plane of \eqref{FLkleingordon} and \eqref{FLtensorkleingordon}, i.e., the zeros of the dielectric functions given above. Specifically, in order to obtain the full spectrum of solutions, one should find, for each value of the wavenumber $k$, the intersection between the curves $\omega_r(k)$ and $\omega_i(k)$ along which the dielectric function vanishes \footnote{We indicate with $\omega_r$ and $\omega_i$ the real and imaginary part of the angular frequency, respectively. The same notation is also used to denote the real and imaginary part of the dielectric functions.}. However, when the weak damping scenario $|\omega_r| \gg |\omega_i|$ is investigated, approximate solutions for the dispersion relation $\omega_r(k)$ and the damping coefficient $\omega_i(k)$ are easily obtained from 
\begin{equation}
    \epsilon^{(\phi,h)}_r(k,\omega_r)=0\label{realpart}
\end{equation}
and
\begin{equation}
    \quad\omega_i=-\left.\frac{ \epsilon^{(\phi,h)}_i(k,\omega)}{\frac{\partial \epsilon^{(\phi,h)}_r(k,\omega)}{\partial\omega}}\right|_{\omega=\omega_r}.
    \label{imaginarypart}
\end{equation}

In particular, we observe that a null imaginary part of the dielectric function corresponds to a purely real angular frequency, leading to the absence of any damping effect. 

On the other hand, by the inspection of \eqref{epsilonscalar} and \eqref{epsilontensor}, we realize that the only imaginary contribution comes from the integration around the Landau pole $\leri{p^0 \omega -k p_3}^{-1}$. Thus, it is clearly demonstrated that a necessary condition for the presence of damping is a subluminal phase velocity $v_p\equiv \frac{\omega_r}{k}<1$; so, that Landau pole is included in the path of integration. Now, we specialize our analysis to the case in which the equilibrium configuration of the medium is described by a J\"uttner distribution

\begin{equation}
    f_0(p)=\dfrac{n}{4 \pi m^2 \Theta K_2 \leri{\frac{m}{\Theta}}}e^{-\frac{\sqrt{m^2+p^2}}{\Theta}},
\end{equation}
where $n$ indicates the number of particles per unit volume, $\Theta$ represents the temperature of the medium expressed in units of the Boltzmann constant $k_B$, and $K_2\leri{\cdot}$ is the modified Bessel function of the second kind. We proceed by inserting the J\"uttner distribution into the expression of the dielectric functions \eqref{epsilonscalar} and \eqref{epsilontensor}: we assume that the phase velocity is much greater than the thermal velocity of the medium; hence, $\frac{\omega}{k}\gg \frac{p_3}{p^0}$ for most of the particles. Under this hypothesis, it is possible to expand the denominators in convergent power series and integrate term by term. Equating to zero, the expressions obtained yield the dispersion relations for the gravitational radiation within the medium. For the tensor modes, it is found that

\begin{equation}\label{disp1}
    \omega_r^2(k)=\frac{k^2+12\omega^2_h\frac{x-\gamma(x)}{x^2}+\sqrt{\leri{k^2+12\omega_h^2\frac{x-\gamma(x)}{x^2}}^2+48k^2\omega_h^2\frac{\gamma(x)}{x^2}}}{2},
\end{equation}
with $x\equiv \frac{m}{\Theta}$ being the ratio between rest and thermal energy, and $\gamma(x)\equiv\frac{K_1(x)}{K_2(x)}$ and $\omega_h^2\equiv\frac{\kappa'' n m}{6}$ being the squared proper frequency of the medium for tensor modes 
(see~\eqref{omegapropria} for a comparison). A simple calculation shows that the phase velocity for tensor gravitational waves is always greater than unity throughout the propagation in the material medium, hence, resulting in a null imaginary part for the angular frequency and in the absence of damping, in agreement with the literature on the theme. Conversely, repeating the same steps for the dielectric function relative to the scalar mode yields
\begin{equation}
       \omega_r^2(k)=\dfrac{k^2+M^2-9\gamma\omega_0^2 +\sqrt{\leri{k^2+M^2-9\gamma\omega_0^2}^2+12\gamma\omega_0^2 k^2}}{2},
   \label{realpartomega}
\end{equation}
where, in this case, $\omega_0$ represents the proper frequency of the medium for scalar waves, i.e., $\omega_0^2\equiv\frac{b \kappa' n m}{6}$. Now, in order to have a subluminal phase velocity, it results that the following inequality must hold
\begin{equation}
    M^2<6\gamma\omega_0^2.
    \label{conditionfordamping}
\end{equation}

This condition represents a distinctive trait of the gravitational Landau damping, which is absent in the electromagnetic counterpart. By relating the mass of the scalar wave, which depends ultimately on the parameters of the gravitational theory taken into account, to quantities describing physical features of the ensemble of particles, it selects media that are able to damp massive modes from noninteracting ones. We point out, moreover, that the constraints on the speed of propagation of gravitational waves derived in~\cite{LIGOScientific:2017vwq,LIGOScientific:2017zic} are translated in our case to a maximum allowed value for the mass $M$ of the scalar degree of freedom. However, such a result does not represent a limitation for our findings, considering that for decreasing values of the mass of the scalar mode, the set of material media able to induce damping becomes actually larger, as it can be easily observed from \eqref{conditionfordamping}. In addition to this, it is worth outlining that the limit of vanishing mass does not correspond to general relativity, since, in this case, we actually deal with a Horndeski theory endowed with a  purely transverse scalar wave, which acts as a breathing in the orthogonal plane. In this respect, the pursued analysis remains strictly valid: for such a massless scalar mode, indeed, Landau damping always occurs, as \eqref{conditionfordamping} is 
identically satisfied by any medium.
Another interesting aspect of inequality \eqref{conditionfordamping} is disclosed if we apply our model to cosmology. Indeed, it can be observed that the quantity $\gamma \omega_0^2$ is always decreasing for decreasing redshift. Hence, if we assume that the mass of the scalar gravitational wave $M$ does not change with time, then for each specific model of material medium there exists a single redshift (at least negative) below which inequality \eqref{conditionfordamping} is no longer satisfied, so that the medium considered becomes unable to damp the scalar radiation. 
We evaluate the imaginary part of the dielectric function \eqref{epsilonscalar} by exploiting the residue theorem; then, we obtain, from \eqref{imaginarypart}, the following form of the damping rate 

\begin{equation}\label{immaginarypartomega}
    \omega_i(k)=-\frac{\pi x}{4 k K_1(x) }\frac{\omega_r^4(k^2+M^2-\omega_r^2)e^{-\frac{x}{\sqrt{1-\frac{\omega_r^2}{k^2}}}}}{3\omega_r^4-2\omega_r^2+(k^2+M^2)}.
\end{equation}

\begin{figure} [hbtp]
    \includegraphics[width=.93\columnwidth]{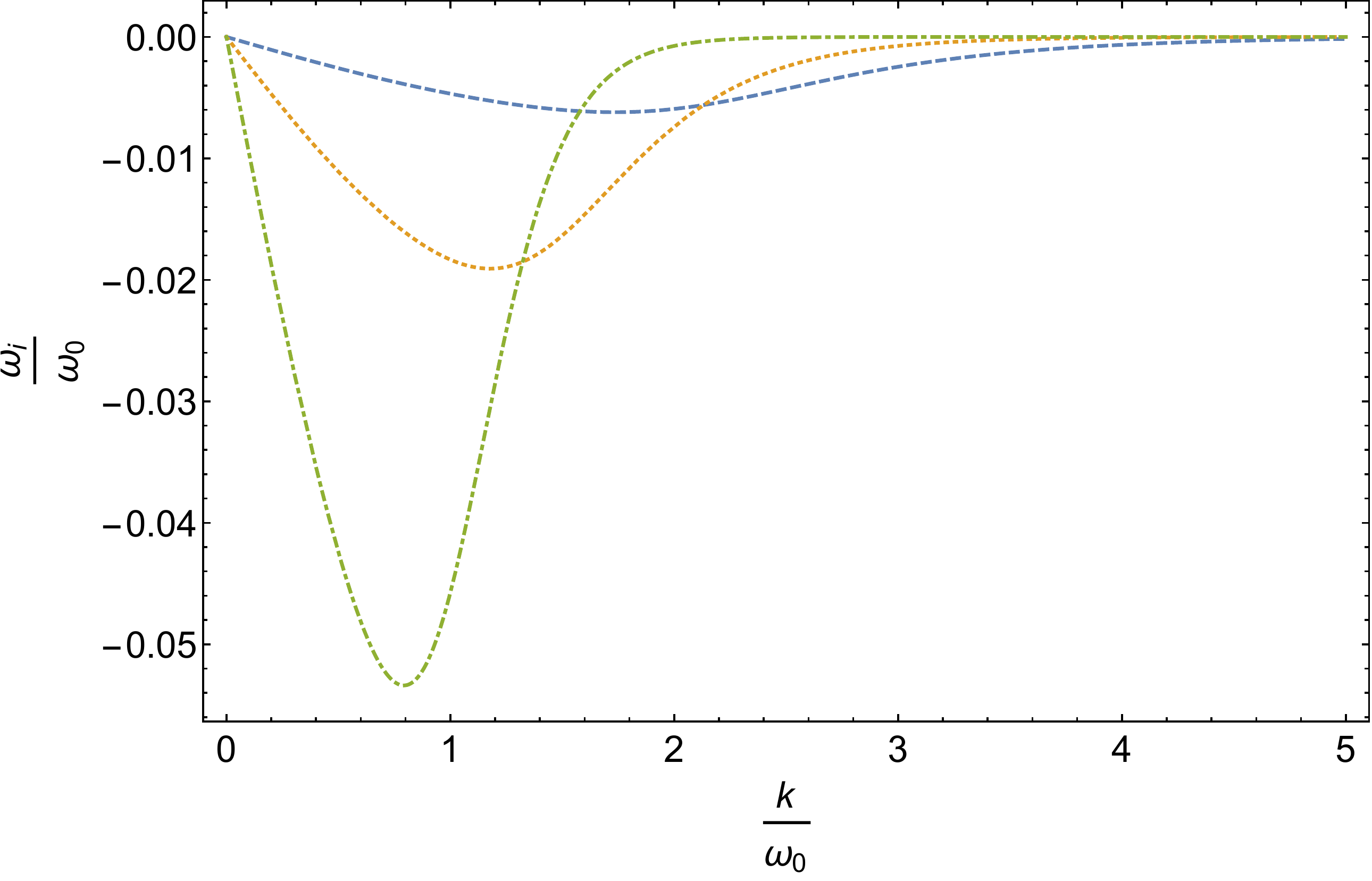}
    \caption{Damping coefficient $\bar{\omega}_i$  vs.  wavenumber $\bar{k}$ for $x=1$ (dashed curve), $x=2$ (dotted curve), and $x=5$ (dot-dashed curve).}
    \label{fig1}
\end{figure}

\begin{figure}[hbtp]
    \includegraphics[width=.93\columnwidth]{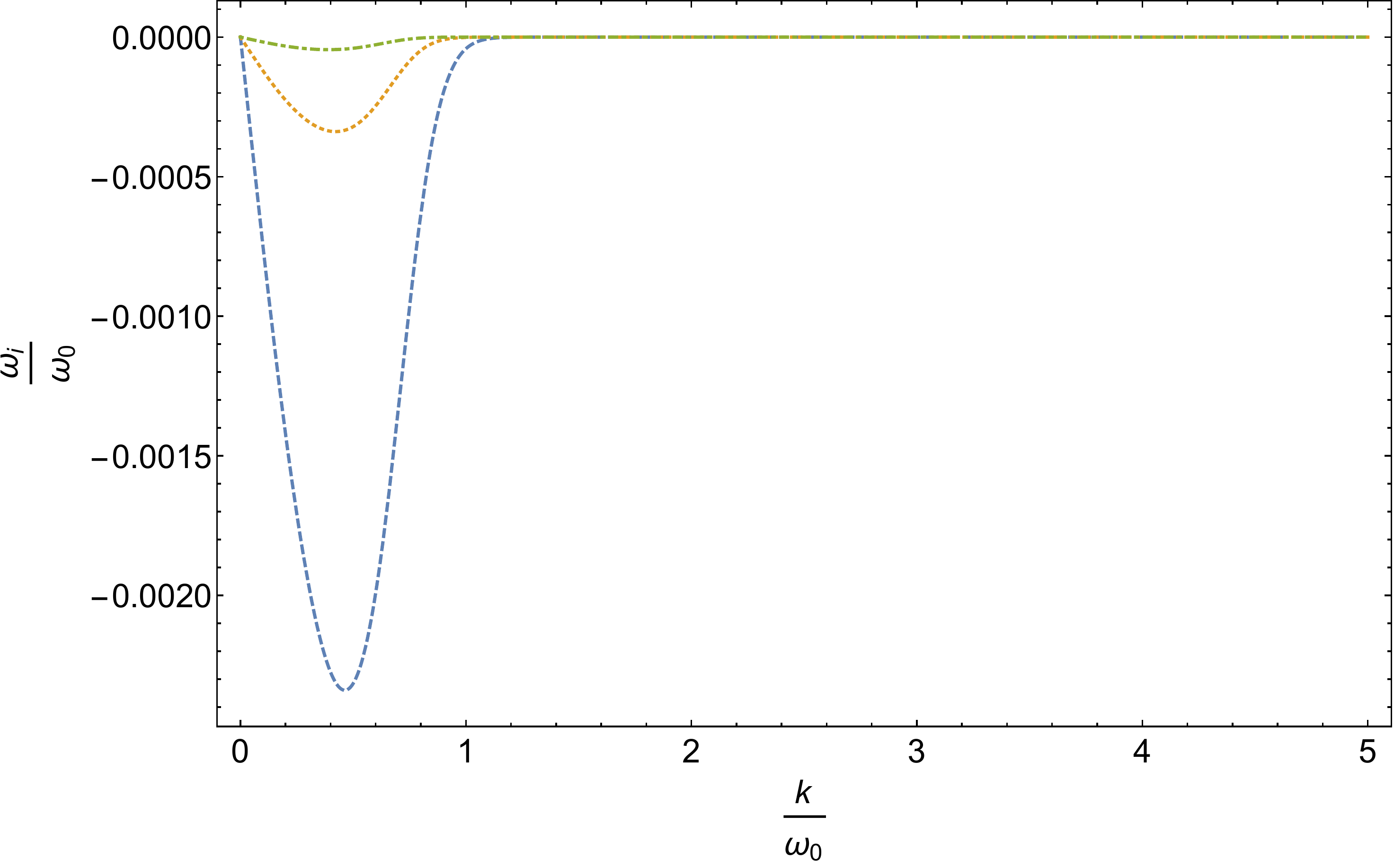}
    \caption{Damping coefficient $\bar{\omega}_i$  vs.  wavenumber $\bar{k}$ for $x=30$ (dashed curve), $x=40$ (dotted curve), and $x=50$ (dot-dashed curve). }
    \label{fig2}
\end{figure}

When the dispersion relation \eqref{realpartomega} is inserted into this expression it can be easily shown that $\omega_i$ results negative for any value of the wavenumber. This is certainly not a surprising fact, as we know from the electromagnetic Landau damping theory that the sign of the imaginary part of the angular frequency coincides with the sign of the derivative of the background distribution function considered at a velocity equal to the phase velocity of the signal. Therefore, given that we have pursued our analysis by assuming a phase velocity of the wave in the fast-tail region of the distribution, an always-negative damping rate is naturally obtained. However, the aforementioned ansatz is not feasible when highly relativistic media are considered: in fact, it can be shown that the J\"uttner distribution is characterized by a thermal velocity very close to the speed of light when the value of the parameter $x$ drops below unity. In such cases, it is required a numerical treatment of the integral contained in the expression of the dielectric function. A simple integration with rectangles and the expansion of the expression obtained in terms of the parameter $\delta \equiv 1- \frac{\omega_r}{k}$, assumed to be small and positive up to third order, yields a family of curves~\footnote{We denote with a bar quantities that are normalized with the proper frequency $\omega_0$, e.g., $\bar{\omega}_i\equiv \frac{\omega_i}{\omega_0}$ and equivalently for the others.}  $\bar{\omega}_i=\bar{\omega}_i \leri{\bar{k}; x,\bar{m}}$ shows relevant similarities with \eqref{immaginarypartomega}, such as, for instance, being always negative with a minimum located roughly around $\bar{k} \simeq \bar{M}$. We report the graphics of the curves obtained by varying the parameter $x$ while keeping the normalized mass fixed $\bar{M}=1$. In particular, in Figure~\ref{fig1}, it can be noticed that the value of the maximum damping grows with $x$ until the value $x\simeq 5$ is reached. Then, as is clearly shown in Figure~\ref{fig2}, the effect rapidly vanishes as soon as $x$ becomes larger than a few tens. However, in both figures, we observe that damping occurs only in a narrow region of $\bar{k}$: in fact, the damping coefficient results are negligible when wavenumbers slightly larger than $\bar{M}$ are considered. This finding suggests some sort of resemblance with spectral lines, giving rise to the introduction of the concept of gravitational-wave spectroscopy. Indeed, by estimating the amount of damping suffered by perturbations of a certain size and by assuming a definite model of modified gravity, it is possible, in principle, to measure the physical properties of the medium traversed by the gravitational radiation. Conversely, by taking into account a precise model of material medium i.e., by setting the mass of the particles, their volume density, and the temperature it is possible to give a quantitative estimate of the damping suffered by scalar waves of a certain spatial size. For instance, considering the interaction between the cosmological dark matter medium and scalar radiation at a redshift $z \approx 2000$, we calculate from our formulae an estimated maximum damping of roughly $\omega_i \lesssim 10^{-16}\,
\text{Hz}  $ (see~\cite{Moretti:2020kpp} for details). It is interesting to notice that in~\cite{Lu:2018smr}, a similar value for the damping rate was directly inferred from observational data, i.e., $\omega_i \lesssim 10^{-17}\,
\text{Hz}  $. The interpretation of such a measurement requires, in the context of general relativity, the introduction of a nonzero viscosity for the medium traversed 
(see Section~\ref{sec3}), so that the measurement of the damping allows the indirect estimate of the viscosity of the Universe. On the other hand, given the value of damping rate obtained in our model, the same data analysis can be reinterpreted within the broader theoretical scenario of modified theories of gravity as the result of kinetic damping suffered by scalar radiation in the interaction with viable models of dark matter. Another important occurrence caused by the possibility of Landau damping for scalar gravitational waves is that the interaction between radiation and material medium tends, on average, to heat up the latter. Then, if we consider scalar modes propagating into the cosmological dark matter medium in the early Universe, we expect that the energy exchange allowed by the phenomenon we have described should result in a global heating of dark matter. This latter fact could have significant observational implications, such as a modification in the time for a structure formation through Jeans instability. However, the linear treatment enforced in our analysis is not sufficient to rigorously quantify the energy exchange rate between radiation and material medium because, in this case, the role of the dominant nonlinear terms in Vlasov equation becomes relevant and the theoretical analysis must be deeply~revised.       

\section{Concluding Remarks}\label{sec4}
In this review, we performed an interesting analogy between properties of photons in electromagnetic plasmas and longitudinal modes of gravitational waves in modified theories of gravity. In particular, we showed how longitudinal excitations are, in general, related to massive states of the graviton, both when they are carried by additional degrees of freedom explicitly introduced in the Lagrangian and when they emerge at the effective level as the result of some averaging procedure of the matter source. This motivated us to trace a comparison with the behavior of photons in ordinary plasmas, where it is a well-established fact that the interaction of electromagnetic waves with the ions can lead to the appearance of longitudinal modes, attributable to the presence of effective massive photons in the medium.

Firstly, we offered a rather exhaustive discussion about the nature and the propagation of massive scalar and vector modes in Horndeski gravity by presenting a gauge-invariant formulation; this permitted us to unequivocally identify the radiating degrees of freedom, thus, allowing 
us to study the phenomenological signature of the polarizations via the geodesic deviation equation. In this respect, besides the ordinary tensor modes, we clearly identified scalar and the vector components, responsible for breathing, longitudinal, and vector polarizations. This very general conclusion provided a precise hint towards the direction that the new generation of interferometer devices could be pushed in order to have a chance to see modified gravity physics.

We focused, then, on the interaction between the gravitational radiation and the traveled medium, and determined three different scenarios characterized by peculiar phenomenology. We first discussed the case of a standard gravitational wave propagating in the presence of matter, with the aim of elucidating how the damping of the amplitude can settle down due to the dissipation properties of the medium, or to the redshift associated to the Universe expansion (particular attention was dedicated to the gravitational \mbox{wave--neutrino} thermal background interaction). In this respect, we outlined that an ordinary gravitational wave, propagating inside an homogeneous and isotropic medium, is not affected by damping mechanism in a tangible manner. We revised the subtle question concerning the propagation of ordinary gravitational waves within molecular matter, where, due to the presence of bounded substructures, we are forced to abandon a continuum representation for the medium. It was immediate to recognize that such a situation is often well-implemented in astrophysical and cosmological systems, in which different spatial scales of gravitationally bounded systems are naturally present. This approach was summarized by the construction of a Macroscopic Theory of Gravity, according to which the deformation of the molecular medium can affect the gravitational radiation and give rise at the effective level to a massive graviton. Indeed, in close analogy with the propagation of electromagnetic waves within a real plasma, the molecule backreaction can be described by the emergence of additional modes able to induce longitudinal stresses on test particles. However, it should be remarked that these extra polarizations are detectable only at the macroscopic spatial scale characterizing the averaging process that defines the macroscopic~theory.

Finally, we discussed the most recent and maybe innovative phenomenon, concerning the Landau damping suffered by the massive scalar mode in generic Horndeski theories. We revised in detail how, even when the matter medium in which the wave propagates is homogeneous and ideal, the scalar mode is suppressed, outlining that the settling of this very specific mechanism depends crucially on a phenomenological inequality relating to the thermodynamic properties of the medium and the mass of the scalar mode.
The analysis followed a kinetic approach in which a noncollisional medium, described by the Vlasov equation, is dynamically coupled to the scalar degree of freedom of Horndeski gravity, and the resulting interaction is expressed in terms of the (gravitational) dielectric properties of the medium (we just considered a linear backreaction of the wave on the~particles). 

This result opens a new interesting perspective on the generality of such a Landau damping within the context of modified gravity approaches, where further scalar and vector degrees of freedom are taken into account. Indeed, as the analysis carried out in Section \ref{sec3.2} clearly shows, the longitudinal character of a specific mode is not a necessary condition for the settling of the damping, which rather turns out to depend on the peculiar dispersion relation of the mode in the medium traversed. Landau damping occurs only if the latter allows for a subluminal phase velocity of the considered radiating degree of freedom, at least in some regions of the wavelengths spectrum. In this respect, an intriguing possibility is offered by further extensions of Horndeski theories, such as the teleparallel formulation~\cite{Bahamonde:2019ipm,Bahamonde:2021dqn} or the Palatini generalization~\cite{Dong:2021jtd}, exhibiting a large variety of additional polarizations, both massive and massless, which could be, in principle, subject to kinetic damping phenomena. 

We conclude by stressing that the achievement of this very result, i.e., the gravitational Landau damping, was based on a very intriguing point of view, according to which the concept of gravitational plasma can be introduced. A crucial step in this direction must be considered the identification of inertial forces as the neutralizing background, analogous to the role played by cold ions in a electromagnetic plasma. This idea, implicitly formulated in~\cite{1962MNRAS.124..279L}, permitted~\cite{Moretti:2020kpp} to study the self-consistent oscillations inside a gravitational medium that is, the 
gravitational equivalent to Langmuir waves in standard plasmas.

\bibliography{references.bib}
\end{document}